\documentclass[aps,prb,twocolumn,draft,showpacs]{revtex4}

\begin{document}

\title{Variational calculation of many-body wave functions and energies
from density-functional theory}
\author{K. Capelle}
\email{capelle@iqsc.sc.usp.br}
\affiliation{Departamento de Qu\'{\i}mica e F\'{\i}sica Molecular\\
Instituto de Qu\'{\i}mica de S\~ao Carlos\\
Universidade de S\~ao Paulo\\
Caixa Postal 780, S\~ao Carlos, 13560-970 SP, Brazil}

\date{\today}

\begin{abstract}
A generating coordinate is introduced into the exchange-correlation functional
of density-functional theory (DFT). The many-body wave function is represented 
as a superposition of Kohn-Sham (KS) Slater determinants arising from different
values of the generating coordinate. This superposition is used to 
variationally calculate many-body energies and wave functions from solutions 
of the KS equation of DFT. The method works for ground and excited 
states, and does not depend on identifying the KS orbitals and energies 
with physical ones. Numerical application to the Helium isoelectronic series 
illustrates the method's viability and potential.
\end{abstract}


\maketitle

\newcommand{\be}{\begin{equation}}
\newcommand{\ee}{\end{equation}}
\newcommand{\bi}{\bibitem}
\newcommand{\la}{\langle}
\newcommand{\ra}{\rangle}

Density-functional theory (DFT) is routinely used in quantum chemistry and
physics for the calculation of, e.g., ground-state energies and charge 
distributions. 
In a typical application of DFT one solves the Kohn-Sham (KS) single-particle
equation, with a suitable choice for the exchange-correlation functional,
and uses the resulting single-particle orbitals to construct the ground-state
particle density. From this density other observables, such as the ground-state
energy, can then be calculated. It is well known that the KS Slater determinant
is not in itself an approximation to the many-body wave function but only
a device for reproducing the correct density. Similarly, the single-particle
eigenvalues of the KS equation do not represent the energy spectrum of the 
system under study, and their differences are not excitation energies.
Although excited-state energies and many-body wave functions for both ground 
and excited states are, in principle, functionals of the ground-state 
density,\cite{hk} they are hard to extract from DFT. For excited-state energies
a number of viable methods has been suggested,\cite{ensdft,tddft,nagy,goerling}
but these are more complicated than conventional ground-state DFT, and not yet 
as widely used. Many-body wave functions cannot be calculated at all from 
standard KS-type calculations.

In the present paper a novel approach based on DFT is proposed, which allows
to (i) systematically improve the accuracy of ground-state energies obtained 
in a standard KS calculation, (ii) calculate excited-state energies, and
(iii) obtain variational approximations for the many-body wave functions
corresponding to these energies.
The device which makes this possible is the Griffin-Hill-Wheeler (GHW) 
variational method,\cite{wheeler,hill} or generator-coordinate method (GCM), 
which is now briefly described. 
In the GHW approach one writes the trial wave function $\Psi$ as an integral 
transformation of a generating wave function $\Phi$, according to
\be
\Psi = \int d\alpha\, f(\alpha) \Phi(\alpha),
\label{trial}
\ee
where $\alpha$ is a generating coordinate and
the variational principle is applied to the kernel of the integral,
$f(\alpha)$. The nature of $\Psi$ and $\Phi$ (i.e., whether they are many-body
or single-body functions) does not matter at this stage, and their
coordinate arguments have thus been suppressed. Calculation of the expectation
value of the system's Hamiltonian $\hat{H}$ with $\Psi$, followed by 
variation of the result with respect to $f(\alpha)$, yields the so-called GHW 
equation\cite{wheeler}
\be
\int d\alpha'\,\left[K(\alpha,\alpha')-ES(\alpha,\alpha') \right] 
f(\alpha') =0.
\label{ghw}
\ee
Here 
$K(\alpha,\alpha')= \la \Phi(\alpha)| \hat{H} | \Phi(\alpha') \ra $
and
$S(\alpha,\alpha')= \la \Phi(\alpha)| \Phi_(\alpha') \ra $ are matrix elements 
of the Hamiltonian with the generator functions $\Phi(\alpha)$, and the 
overlap of these functions, respectively. Solution of the GHW equation
(\ref{ghw}) yields the energies $E$, which are variational approximations 
for the eigenvalues of $\hat{H}$, and the function $f(\alpha)$ which through
(\ref{trial}) determines the corresponding wave functions. 
The full set of solutions of the eigenvalue problem (\ref{ghw}) yields
thus the spectrum and wave functions of Hamiltonian $\hat{H}$.

Originally this method arose in nuclear physics,\cite{wheeler,hill} where 
$\Psi$ was taken to be the nuclear many-body wave function, and the
generating coordinate $\alpha$ was interpreted as a `deformation parameter'
describing collective oscillations of the nucleus. The generator 
functions $\Phi(\alpha)$ were obtained from solving an auxiliary, simplified, 
Schr\"odinger equation in which the nuclear potential $v({\bf r})$ was 
replaced by a deformed potential $v_\alpha({\bf r})$, with $\alpha$ 
characterizing the degree of deformation. Outside nuclear physics the GHW
approach has been applied to various model Hamiltonians,\cite{modelcalcs} 
the electron gas,\cite{providencia} and molecular electronic-structure 
calculations,\cite{laskowski1,laskowski2,laskowski3} exploring a variety of
different choices for the generator functions $\Phi$ and the generating 
coordinate $\alpha$.
More recently, it has been used to generate optimized basis functions for 
Hartree-Fock calculations, starting from a set of simple trial 
functions.\cite{ghwhf1,ghwhf2,ghwhf3} In this type of application 
$\alpha$ is identified with the basis function exponent $\zeta$, and $\Psi$ 
and $\Phi$ are single-particle functions.

In the present contribution we go back to many-body wave functions. The basic
idea is to identify the auxiliary Schr\"odinger equation with the KS equation 
of DFT, the deformation potential producing the family of generator functions
with the KS potential, and the deformation parameter (or generating 
coordinate) with a parameter in the exchange-correlation ($xc$)
functional. Many functionals, such as the $X\alpha$ and B3-LYP
approximations, naturally contain such a parameter, but it can always
be introduced by hand in any functional. 
The present proposal is thus to write the many-body wave function as
\be
\Psi({\bf r}_1,\ldots{\bf r}_N) = 
\int d\alpha\, f(\alpha) \Phi^{KS}(\alpha;{\bf r}_1,\ldots{\bf r}_N),
\label{kstrial}
\ee 
where $\Phi^{KS}(\alpha)$ is the Slater determinant obtained from a KS 
calculation with $xc$ potential $v_{xc,\alpha}$. 
Note that the deformation parameter $\alpha$ is neither an adjustable parameter 
fixed by comparison with experiment nor a variational parameter for a given 
form of the trial wave function, but rather a generating coordinate 
that accounts for collective behaviour in $\Psi$ not described by the 
single-particle coordinates ${\bf r}_1\ldots {\bf r}_N$ in the Slater 
determinant $\Phi^{KS}(\alpha;{\bf r}_1\ldots {\bf r}_N)$. Its physical origin 
is in the deformations of the single-body potential $v_{xc,\alpha}$, which 
simulate the collective degrees of freedom of the interacting many-body 
system.\cite{wheeler}

The idea proposed here is thus to use DFT potentials and orbitals as input
for a GHW calculation, and the GHW Eq.~(\ref{ghw}) with (\ref{kstrial})
to gain direct variational access to many-body energies and wave functions. 
As a first viability test of this scheme it is now applied
to the Helium isoelectronic series. 
For a general two-electron atom in a closed-shell configuration the
kernels $K$ and $S$ are evaluated easily. $\Phi^{KS}(\alpha)$ is 
a $2\times 2$ Slater determinant formed with one 
doubly occupied orbital. After performing the sum over spins 
one is left with only spatial integrals, which need to be calculated
numerically, and the kernels take the form
$K(\alpha,\alpha')= 2\la \alpha | \alpha'\ra
\la \alpha | \hat{t} + \hat{v} | \alpha'\ra
+ \la \alpha \alpha | \hat{u} | \alpha'\alpha' \ra$,
and $S(\alpha,\alpha')=\la \alpha | \alpha'\ra^2$, where $|\alpha\ra$ 
stands for the KS orbital $\varphi_\alpha({\bf r})$, and
$\hat{t}$ and $\hat{v}$ are the single-particle kinetic and potential
energy operators. $\hat{u}=1/|{\bf r}-{\bf r}'|$.
Note that the diagonal element ($\alpha=\alpha'$) of the kernel $K$
is simply the energy expression one obtains in a restricted closed-shell
Hartree-Fock calculation\cite{szabo} for a two-electron atom. A similar
formal connection to Hartree-Fock theory will always hold, since $K$ is 
the matrix element of the Hamiltonian between single Slater determinants.
The single-particle orbital $\varphi_\alpha({\bf r})$ is now obtained from a 
self-consistent KS calculation. For simplicity (and to illustrate an
important point below) I choose as $xc$ functional the $X\alpha$
approximation and take as generating coordinate the 
parameter $\alpha$ present in that functional.
(The fact that the GHW generating coordinate and the $X\alpha$
coefficient are both traditionally called $\alpha$ is a coincidence.)
The GHW integral equation (\ref{ghw}) is then solved by discretization.
A suitable mesh for discretizing the integral over $\alpha$ is
$\{0,0.5,1,1.5,2\}$. Five mesh points may appear surprisingly few,
but empirically it was found that larger or denser meshes did 
not significantly change the results.\cite{footnote1} 
The calculation thus consists in the following three steps:
(i) performing KS $X\alpha$ calculations at the prescribed
values of $\alpha$; (ii) using the resulting orbitals to evaluate the
kernels $K$ and $S$, according to the above equations; and (iii) discretizing 
the integral equation (\ref{ghw}) on this set of $\alpha$'s and solving it
by standard matrix algebra. In the remainder of this paper some representative
results obtained in this way are presented.

\begin{table}
\caption{\label{table1} Ground-state energies of the Helium isoelectronic
series. First row: energies obtained from solving the GHW equation, using the
$X\alpha$ functional to obtain the generator orbitals, as described in
the main text. Second row: exact (nonrelativistic) energies from
Ref.~\protect\onlinecite{froese}. 
Third row: percentual deviation of the GHW energy from the exact one. All 
energies are in atomic units. For comparison: the LDA result for $He$ is 
$E_0^{LDA}=-2.835 a.u.$ and deviates from the exact value by $2.4\%$.}

\begin{ruledtabular}
\begin{tabular}{r|cccccccc}
& $-E_{0}^{GHW-X\alpha}$ & $-E_0^{exact}$ & $\%$ deviation \\
\hline
$He$ & 2.871 & 2.904 & 1.1 \\
$Li^+$ & 7.244 & 7.280 & 0.49 \\
$Be^{2+}$ & 13.62 & 13.66 & 0.29 \\
$B^{3+}$ & 22.00 & 22.03 & 0.14 \\
$C^{4+}$ & 32.37 & 32.41 & 0.12  \\
$N^{5+}$ & 44.75 & 44.78 & 0.067  \\
$O^{6+}$ & 59.12 &  59.16 & 0.068\\
$F^{7+}$ & 75.50 & 75.53 & 0.040
\end{tabular}
\end{ruledtabular}
\end{table}

Table \ref{table1} compares the lowest eigenvalue of Eq. (\ref{ghw}) (i.e., 
our approximation to the many-body ground-state energy) with reference data 
obtained from numerically exact wave functions.\cite{froese} Considering the 
simple discretization scheme and generating functional ($X\alpha$),
the quantitative agreement achieved is rather surprising. The largest 
deviation from the reference data is found for Helium, and is only about 
$1\%$. Interestingly, this agreement has been obtained by
starting out with an $xc$ functional that on its own yields significantly
worse energies: the ground-state energy of $He$ calculated 
from the usual KS scheme, employing the $X\alpha$ functional with the above 
values of $\alpha$, is found to be $E_0^{\alpha=0}=-1.952 a.u.$,
$E_0^{\alpha=0.5}=-2.515 a.u.$, $E_0^{\alpha=1}=-3.170 a.u.$,
$E_0^{\alpha=1.5}=-3.915 a.u.$, and $E_0^{\alpha=2}=-4.749 a.u.$, respectively,
which are all significantly off the true $He$ ground-state energy. 
Out of these unphysical energies the GHW optimization generates
a ground-state energy of $E_0^{GHW-X\alpha}=-2.871 a.u.$, which deviates
only by about one percent from the exact reference value (see Table 
\ref{table1}). 
Since this is a variational calculation it is, of course, rather
natural that the GHW energy is better than that obtained with each generating
functional (i.e., each value of $\alpha$) individually. GHW optimization can 
thus be used to systematically improve on results obtained from a given input 
density functional, which need not be very good on its own.
On the other hand, the $X\alpha$ GHW value $-2.871 a.u.$, found above, is also
closer to the exact result\cite{froese} $E_0^{exact}=-2.904 a.u.$ than the 
ones obtained with Hartree-Fock\cite{froese} ($E_0^{HF}=-2.862 a.u.$) or 
LDA ($E_0^{LDA}=-2.835 a.u.$). This improvement may appear surprising because
the $X\alpha$ functional in itself can be interpreted as a rather simple 
approximation to both Hartree-Fock and LDA, but is simply explained by noting
that in the present context the $X\alpha$ functional (or any other generator
functional from DFT that could be used instead) only serves as a 
convenient way to build a family of continuously parametrized generator
determinants $\Phi(\alpha)$, and is not used directly to obtain the desired 
results.

In principle, the $N$ eigenvalues of the GHW equation found by discretizing
it as an $N\times N$ matrix equation provide the $N$ lowest-lying energies
of the original Hamiltonian, but the quality of the resulting energies will
depend on the nature of the generator functions. The present calculation
takes ground-state KS determinants to form the generator function and thus does
not directly aim at excited states. However, from the second-lowest eigenvalue
of the GHW equation one obtains an estimate for the energy of the lowest
excited state with same symmetry as the ground state.
For $He$ this is the $2^1S_0$ para state, with an energy of
$-2.146 a.u.$. The value found in the above ground-state GHW calculation is
$E_1^{GHW-X\alpha}=-1.788 a.u.$ and deviates from this by $16.7\%$.
A ground-state GHW-DFT calculation thus allows one to obtain
estimates of the energies of excited states with the {\it same symmetry} as
the ground state. It is thus complementary to methods based on minimization
in symmetry subspaces, which give access to excited states of symmetry {\it
different} from the ground state.
Calculations optimized for specific target excited states, using excited-state
KS Slater determinants as generator functions instead of ground-state ones,
are expected to provide better results for the corresponding excitation
energies.

Another interesting application of GHW variational optimization in DFT is the
calculation of many-body wave functions.
Within the GHW scheme the many-body wave function $\Psi$ is known in terms
of the family of generator functions $\Phi(\alpha)$ and the weight function
$f(\alpha)$, by means of Eq.~(\ref{trial}). After discretization, the weight 
function reduces to a set of coefficients $f(\alpha)$, which can be read off 
directly from the components of the eigenvector
corresponding to a given eigenvalue $E$ of Eq.~(\ref{ghw}). These 
coefficients are thus automatically obtained together with the eigenvalues.
As an explicit example, the (unnormalized) many-body wave function 
obtained together with the above result of $E_0^{GHW}=-2.871 a.u.$ for 
the $He$ ground state is
\begin{eqnarray}
\Psi_0^{He}=-0.0529 \Phi_{\alpha=0}^{KS} + 0.276\Phi_{\alpha=0.5}^{KS}
-0.451 \Phi_{\alpha=1}^{KS} 
\nonumber \\
+ 0.770 \Phi_{\alpha=1.5}^{KS} -0.354 \Phi_{\alpha=2}^{KS},
\label{psihe}
\end{eqnarray}
while the one obtained together with $E_0^{GHW}=-59.12 a.u.$ for the
$O^{6+}$ ground state is 
\begin{eqnarray}
\Psi_0^{O^{6+}}=-0.116 \Phi_{\alpha=0}^{KS} + 0.539\Phi_{\alpha=0.5}^{KS}
-0.786\Phi_{\alpha=1}^{KS} 
\nonumber \\ 
+ 0.279 \Phi_{\alpha=1.5}^{KS} + 0.0263 \Phi_{\alpha=2}^{KS},
\label{psio}
\end{eqnarray}
where the $\Phi^{KS}(\alpha)$ are the KS Slater determinants obtained in
the generator $X\alpha$ calculations for $He$ and $O^{6+}$, respectively.
Wave functions for excited states are obtained in just the
same way from the components of the higher eigenvectors. 

In a variational calculation, such as the one performed here, energies are 
typically obtained with higher accuracy than wave functions. However, the 
simple five-term expansions, given above, already suffice to obtain expectation
values of observables that are comparable to those obtained with 
other methods. As an example, consider the expectation value
$\la \Psi_0 | r^n | \Psi_0 \ra$, for $n$ ranging from $-2$ to $+2$. 
Results for the $He$ atom, calculated with the GHW wave 
function (\ref{psihe}) are listed in Table~\ref{table2}. 
In spite of the simplicity of the $X\alpha$ functional and the inferiority of
variational wave functions to variational energies, the expectation values 
obtained with the five-term GHW-optimized $X\alpha$ wave function are found
to be close to those obtained from a much more sophisticated density functional.

\begin{table}
\caption{\label{table2} Expectation value of the operator $r^n$ for various 
values of $n$ calculated with the Helium GHW many-body wave function 
(\ref{psihe}), compared with results from a standard DFT calculation using 
the B88-LYP functional\cite{b88,lyp} (in a.u., and after normalization).}

\begin{ruledtabular}
\begin{tabular}{ccc}
n & $\la r^n\ra_{GHW-X\alpha-DFT}$ & $\la r^n\ra_{B88-LYP-DFT}$\\
\hline
-2 & 5.74 & 5.98 \\
-1 & 1.66 & 1.69 \\
 0 & 1.00 & 1.00 \\
 1 & 0.928 & 0.964 \\ 
 2 & 1.16 & 1.26
\end{tabular}
\end{ruledtabular}
\end{table}

Expressions (\ref{psihe}) and (\ref{psio}) for $\Psi_0$ also allow us to 
see clearly the difference between the present scheme and configuration
interaction (CI). In CI $\Psi_0$ 
is written as a linear combination of Slater determinants that all stem
from the {\it same} HF calculation. Individual determinants differ by 
systematically substituting occupied single-particle orbitals by unoccupied
ones. In the present scheme $\Psi_0$ is also a linear combination of
Slater determinants, but each determinant comes from a {\it different}
KS calculation, the amount and nature of the difference being specified
by the deformation coordinate $\alpha$. The determinants in the present 
expansion thus stem from Hamiltonians with different potentials. This implies 
that each of them can be interpreted as an effective resummation of a large 
number of CI-type determinants arising from a fixed Hamiltonian. 
Another consequence is that the individual determinants in the GHW expansion 
are not mutually orthogonal. Interestingly, representation of 
many-body wave functions in terms of non-orthogonal determinants has previously
been shown, in the context of the resonating Hartree-Fock method, to be an
efficient way to account for strong Coulomb correlations not readily 
accounted for by traditional expansions.\cite{tenno1,tenno2,tenno3}

In summary, the present paper proposes to combine two many-body methods 
(DFT and GHW), each of which is successful in its field of origin 
(electronic-structure theory and nuclear physics, respectively), but 
which had not previously been brought to work together. At the heart of 
the present proposal is Eq.~(\ref{kstrial}), which expresses the many-body
wave function as a weighted superposition of Kohn-Sham determinants, each
arising from a differently deformed exchange-correlation potential. As a 
first viability test the method has been applied to the $He$ isoelectronic 
series. A full judgement of the powers of GHW-DFT must await systematic 
tests for a wide variety of physical systems and generator functionals, but 
the present initial exploration shows that (i) ground-state energies can be
obtained that are considerably better than those calculated from both the 
generator functional and more sophisticated methods approximated by it,
(ii) many-body wave functions are obtained with almost no 
additional numerical effort, and yield expectation values that are close to 
those given by other methods, and (iii) access to excited state energies and 
wave functions is, in principle, possible, too.

{\bf Acknowledgments}
This work was sup\-por\-ted by FAPESP. I thank L.~N.~Oliveira, M.~Teter and 
A.~Niklasson for useful remarks on an earlier version of this paper, and 
M.~Trsic and A.~B.~F.~da~Silva for hospitality at the IQSC.

\end{document}